\documentclass[11pt]{article}

\usepackage[]{graphicx}
\usepackage[squaren]{SIunits}
\usepackage{isolatin1}

\newcommand{\gfrac}[2]{\displaystyle\frac{#1}{#2}}

\newcommand{\dd}{\mbox{d}}

\title{Calorimeter-less gamma-ray telescopes: Optimal measurement of charged particle momentum from multiple scattering by Bayesian analysis of Kalman filtering innovations.}

\author{Denis Bernard ~ \& ~ Mikael Frosini
 \\
 LLR, Ecole Polytechnique, CNRS/IN2P3, 91128 Palaiseau, France
}

\begin{document} 

\maketitle

\begin{center}
\large \textbf{Presented at the 7th Fermi Symposium 2017, 
 Garmisch-Partenkirchen, Germany }
\end{center}

\begin{abstract}
Novel gamma-ray telescope schemes (silicon wafer stacks,
 emulsions, gas detectors) are being developed so
 as to bridge the sensitivity gap between
 Compton and pair-creation telescopes.
The lower average density with respect to the tungsten/silicon
active target of the Fermi-LAT makes large effective-area
telescopes voluminous objects, for which the photon energy measurement
by conventional means (calorimeter, magnetic spectrometer, transition
radiation detector) is a challenge for the mass budget of the space mission.
We present an optimal measurement of track momentum by the multiple
measurement of the angular deflections induced by multiple scattering
in the active target itself, using a Bayesian analysis of the
filtering innovations of a series of Kalman filters applied to the
track.
For a silicon-wafer-stack telescope, the method yields meaningful
results up to a couple of GeV/c.
\end{abstract}


This work originated in the context of a theoretical and experimental development
of low density, homogeneous detectors such as a gas time-projection
chamber as a high-performance gamma-ray telescope and polarimeter in
the $\gamma \rightarrow e^+e^-$ regime, that is, above a photon energy of 
$1\,\mega\electronvolt$
\cite{HARPO}.
In these ``active targets'', the incident $\gamma$ converts and the produced electrons are tracked in the same structure.
The single-track angular resolution of these detectors, if tracking is
performed with an optimal method in the presence of multiple
scattering such as a Kalman filter \cite{Bernard:2012uf}, is so
good that polarimetry has been predicted to be possible despite the
dilution of the 
polarization asymmetry induced by multiple
scattering \cite{Bernard:2013jea}, and has actually been
demonstrated by the characterization of a TPC prototype on
beam \cite{Gros:SPIE:2016,Gros:2017wyj}.
Other projects plan to use all-silicon active targets, that is,
without any additional tungsten converters
\cite{E-Astrogam:2016,AMEGO}.
A third possibility is the use of emulsion active targets
\cite{Takahashi:2015jza}.
Whatever the method used, the lower density and the lower average
atomic number of the active target, with respect to the tracker of the Fermi-LAT,
imply a larger volume at a given value of the effective area:
measuring the photon energy, be it with a calorimeter or by a
measurement of the electron momenta with a magnetic spectrometer or
with a transition radiation detector, is a challenge to the mission
mass budget.

Lower energies are the domain where Compton telescopes are most
efficient.
When one scatter is detected, from the measurement of the position
and the energy of the recoil electron and of the absorbed scattered
photon, the reconstruction of the direction of the incident photon
constrains it to a large arc, that is a large point spread function
(psf).
If in addition the direction of the recoil electron is
measured, the extension of the arc is reduced and the psf is
improved \cite{Tanimori:2017ihu}.
Alas, the angular coverage of the calorimeter is an issue: it must be
low enough that the incident photon can enter the detector and large
enough to measure the energy of the scattered photon: in this
electron-tracking Compton camera (ETCC) scheme the effective area
undergoes a sharp drop for photon energies above
$0.5\,\mega\electronvolt$ \cite{Tanimori:2017ihu}.

The measurement of the angle deflections induced by 
multiple scattering of the electrons in the tracker has 
been used to yield an estimate of the track momentum since the 1950's 
\cite{Moliere}.
Multiple scattering expresses the combined angle deflection induced by the multiple single deflections of a charged particle due to its many close encounters with the ions and with the electrons
of matter as a Gaussian distribution with RMS 
\begin{equation}
\theta_0 \approx \gfrac{p_0}{\beta p} 
\sqrt{\gfrac{\Delta x}{X_0}} \left( 1 + \epsilon \log{\gfrac{\Delta x}{X_0}}\right),
 \label{eq:multiple:scattering:base}
\end{equation}
where $p_0 = 13.6\,\mega\electronvolt/c$ is the
multiple-scattering constant, $\Delta x$ is the matter thickness through
which the particle propagates and $X_0$ is its radiation length
 \cite{Olive:2016xmw}.
We neglect the small logarithmic correction term parametrized by
 $\epsilon$.

As $\theta_0$ is inversely proportional to the track momentum $p$ and
as each deflection measurement $\theta$ can be considered as an
estimate of the RMS $\theta_0$, an estimate of the momentum can be obtained 
from multiple measurements of $\theta$ along the track
\cite{Moliere}.
Tracks are usually segmented into tracklets, each of which is fit so
as to measure its angle.
The comparison of the angles of successive tracklets yields the
measurement of the deflection.
The choice of the detector segmentation is of utmost importance in
particular in the case of detectors that are physically segmented such
as for silicon detectors.
Shorter segments enable more deflection measurements but each of which with a
larger angle uncertainty.
A back-of-the-envelope guestimate of the optimal value of the tracklet
length $\Delta$ was obtained \cite{Bernard:2012uf},
assuming that the expression of the relative momentum uncertainty
$\sigma_p/p$ is the sum of the low-$\Delta$ and of the
high-$\Delta$ asymptotes.
It was found that the optimal segmentation depends on the value of the
track momentum that we aim to measure.

Kalman-filter-based tracking is known to take into account the
single-point spatial precision of the tracker and multiple scattering
in an optimal way, but only at the condition that the ``noise''
covariance matrices, that in particular of the momentum-dependent
multiple-scattering ``process'' noise, be known
\cite{Fruhwirth:1987fm}.

How devise a segmentation-free, optimal, unbiased, momentum
measurement from the amazing ability of Kalman filters to
(statistically) decipher the (detector precision) uncorrelated and the
(multiple scattering) correlated interleaved contributions in the
track trajectory statistics ?
We develop an optimal method to measure the electron momentum from the
multiple measurement of multiple scattering induced deflections, based
on a Bayesian analysis of the innovation residues of a set of
momentum-dependent Kalman filters applied to the track
\cite{Frosini:2017ftq}.
\footnote{This work has been performed under a number of assumptions/approximations.
We assume relativistic particles ($\beta \approx 1$) without loss of generality.
Only the first-order term (angle deflection) of multiple scattering is
taken into account which is legitimate for the thin detectors
considered here;
the 2nd-order transverse displacement is neglected. 
Continuous ($\dd E / \dd x$) and discrete (BremsStrahlung radiation) 
energy losses are also neglected.
In TPCs in which the signal is sampled, most often the electronics
applies a shaping of the pulse before digitisation, that creates a
short-scale longitudinal correlation between successive measurements
that we neglect too.
Also the limitations of pattern recognition, that is, in the case of
pair
$\gamma$-ray telescopes, of the assignment of each hit to one of two
close tracks, are not addressed.}
Let's define $s$, the track-momentum-dependent 
multiple-scattering-angle average variance per unit track length:
\begin{equation}
s \equiv \left(\gfrac{p_0}{p}\right)^2 \gfrac{\Delta x}{l X_0}
.
\end{equation}

It can be shown by the recurrence application of the Bayes theorem
\cite{Matisko:Havlena:2013} that the probability
$p_n(s) = p_n(z_0 \cdots z_n | s)$ of observing the $n+1$ first
measurements along the track, $z_0 \cdots z_n$, given $s$, can be
expressed as the product of the probability density functions of the
innovations of a Kalman-filter applied to $z_0 \cdots z_n$.
$l$ is the longitudinal sampling (in the case of a homogeneous
detector, the thickness of the scatterer is equal to the length of the
longitudinal sampling, $l = \Delta x$).
\begin{figure}[h]
\centering 
\includegraphics[width=0.475\textwidth]{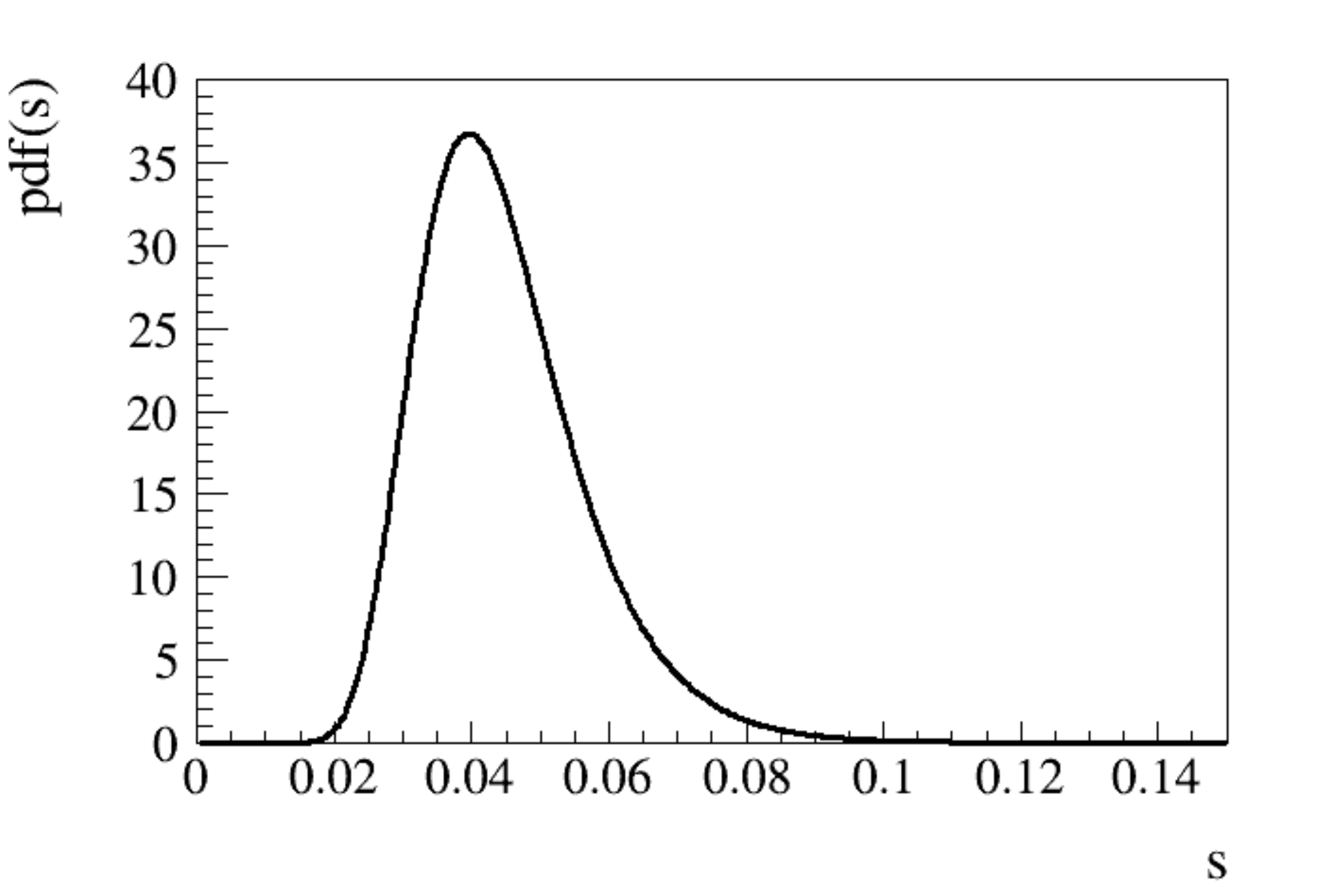}
\hfill
\includegraphics[width=0.475\textwidth]{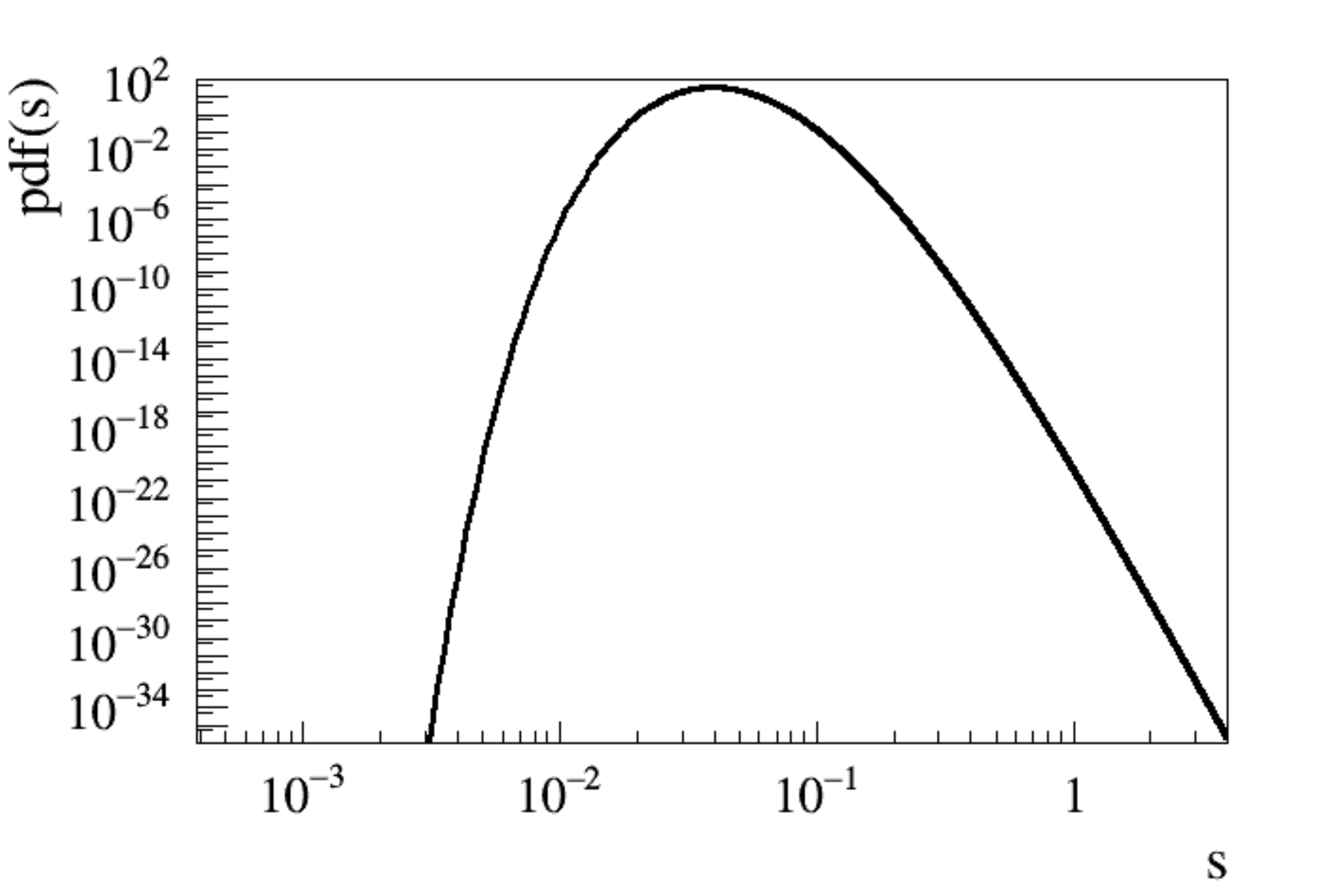}
\caption{\label{fig:p:s}$p(s)$ distribution for a $50\,\mega\electronvolt/c$ track in a silicon detector.
 On that track, the momentum is measured to be equal to
 $49.9\,\mega\electronvolt/c$
\cite{Frosini:2017ftq} (left: lin-lin; right: log-log).
}
\end{figure}

The distribution of the probability density function $p(s)$ is shown
in Fig. \ref{fig:p:s} for one simulated $50\,\mega\electronvolt/c$
track \cite{Frosini:2017ftq}.
The most probable value of $p$ is then obtained from the most probable
value of $s$:
$ p = p_0 \sqrt{{\Delta x}/{l X_0 s}}$.

\begin{figure}[h]
\includegraphics[width=0.45\linewidth]{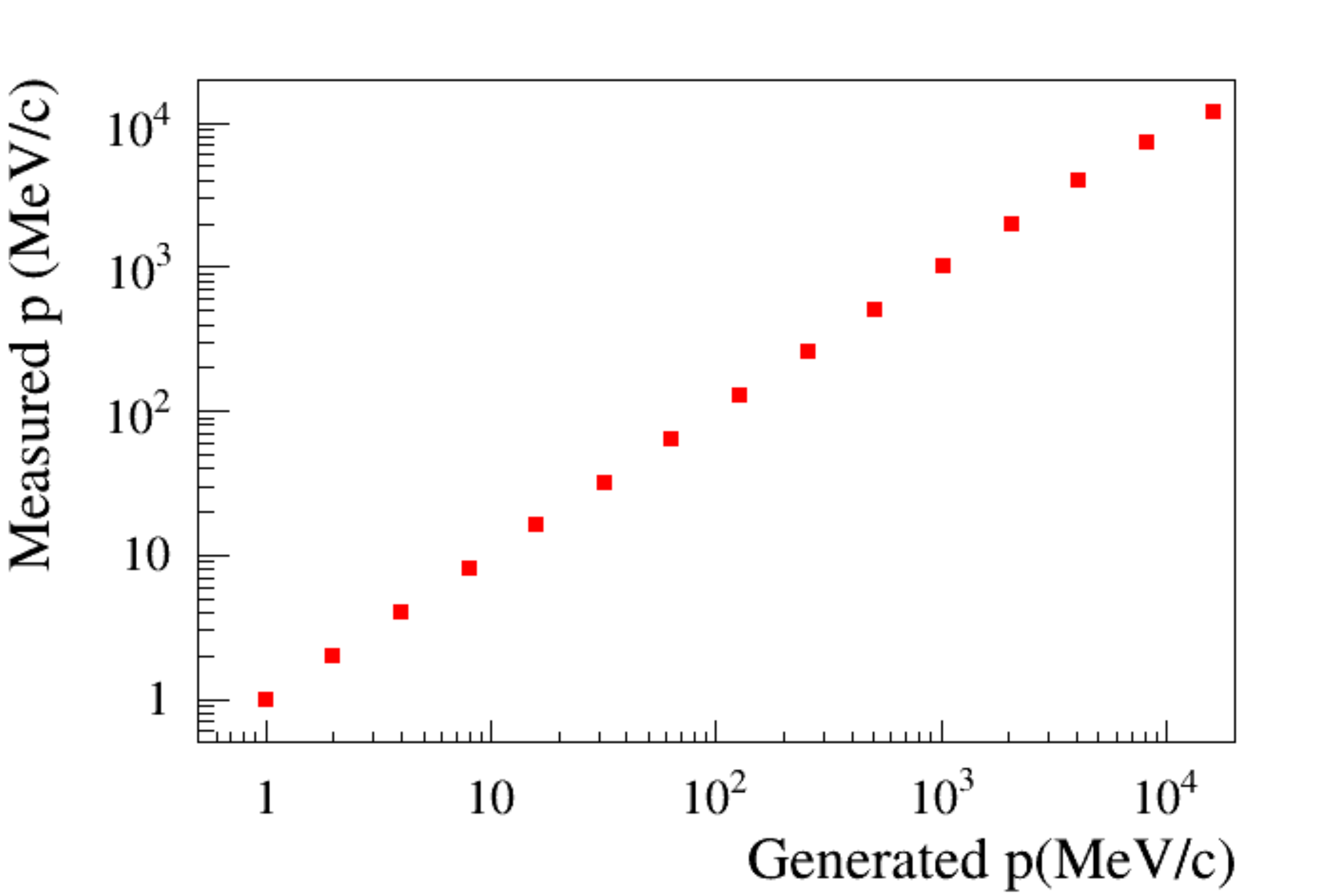}
\includegraphics[width=0.45\linewidth]{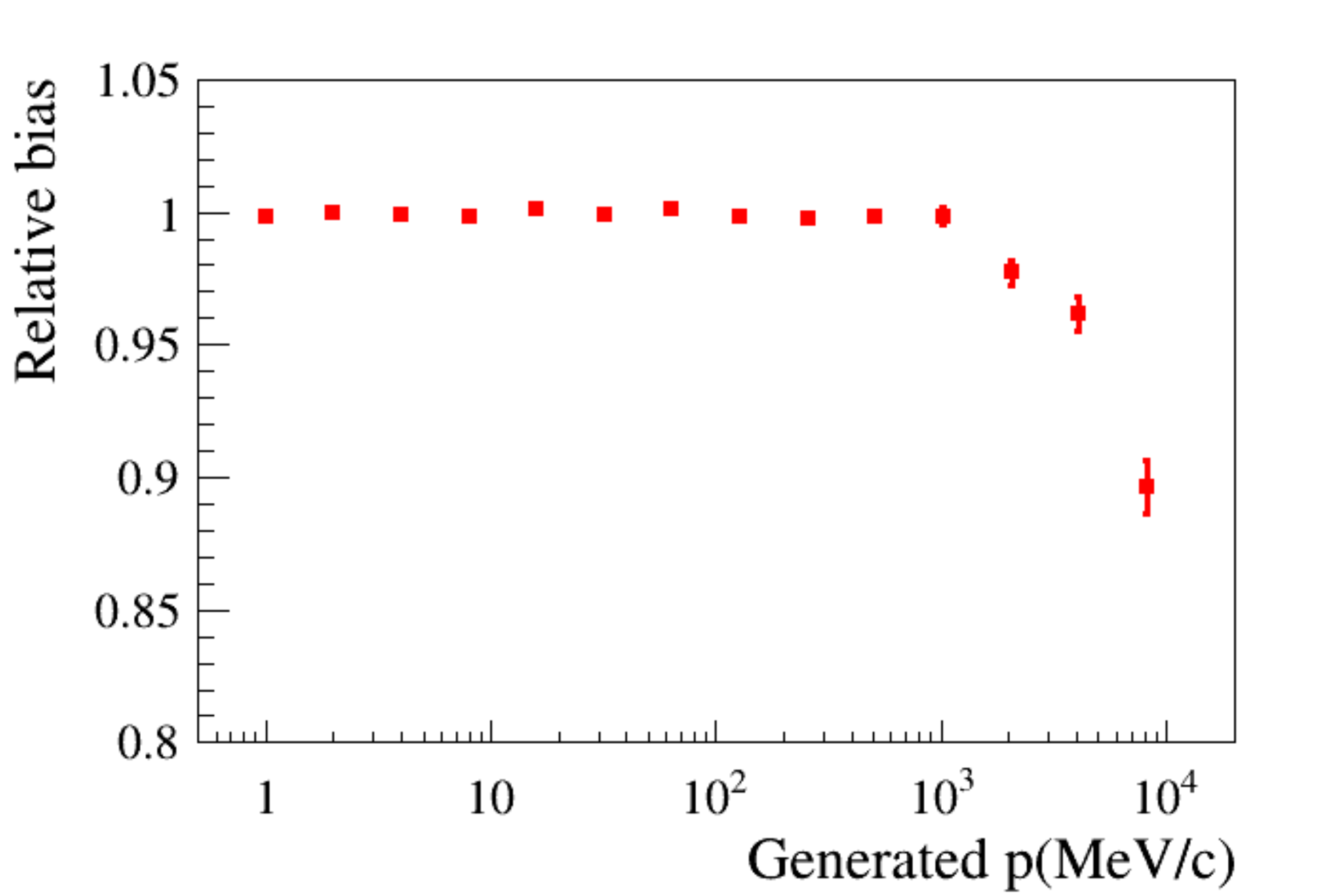}
 \put (-330,80) {(a)}
 \put (-130,80) {(b)}

\includegraphics[width=0.45\linewidth]{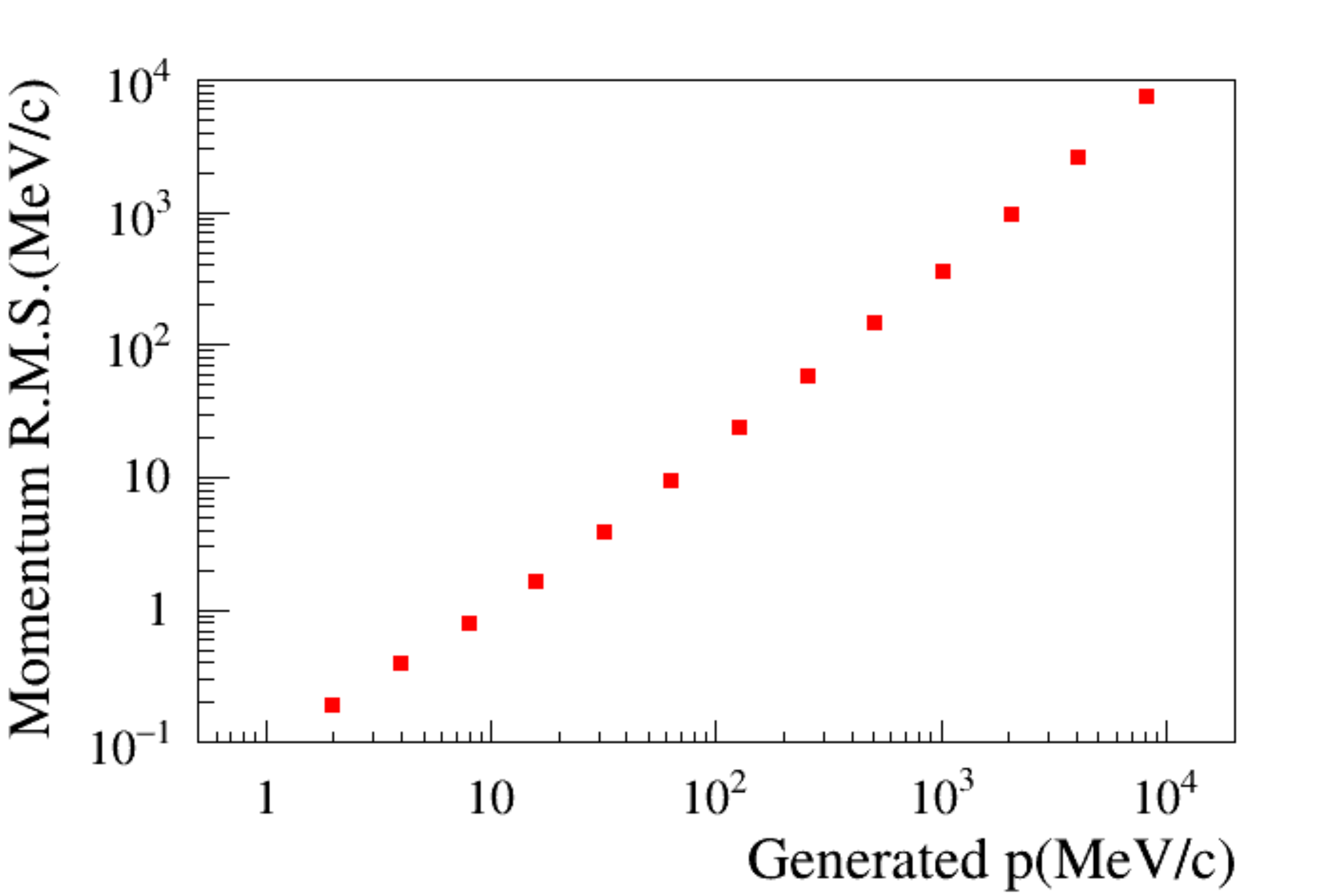}
\includegraphics[width=0.45\linewidth]{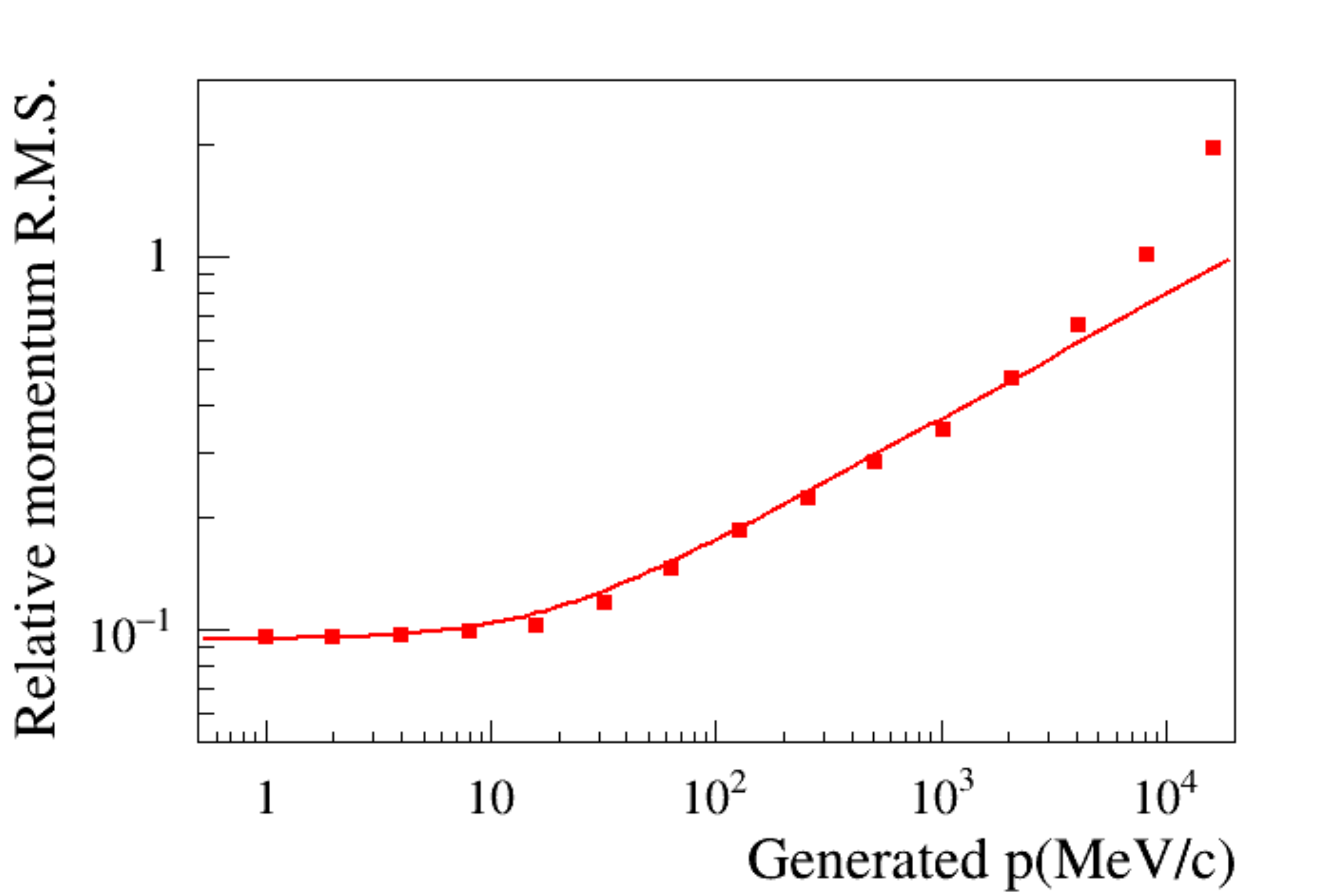}
 \put (-330,80) {(c)}
 \put (-150,80) {(d)}

\caption{\label{fig:silicium}
 Performance of the momentum measurement for the silicon detector: 
 Variation as a function of the true (generated) particle momentum of
 (a) the average measured momentum;
 (b) the average measured normalized to the generated momentum;
 (c) R.M.S of the measured momenta;
 (d) the relative R.M.S of the measured momenta
\cite{Frosini:2017ftq}.
The curve is from eq. (\ref{eq:sigma:sur:p:parametrique}).
For each momentum value, the calculation is based on a
sample of $10^{4}$ simulated tracks.}
\end{figure}

The method provides an optimal measurement of the process-noise parameters 
\cite{Matisko:Havlena:2013}. 
A numerical characterisation of the method shows that for a given
detector the method is reliable up to some limit momentum 
above which the relative precision $\sigma_p / p$ becomes larger than unity.
For lower momentum tracks, the momentum estimation is
found to be unbiased (Fig. \ref{fig:silicium}).
The method is found to be usable at low momentum, below a couple of 
$\giga\electronvolt/c$ for silicon detectors such as e-ASTROGAM
\cite{E-Astrogam:2016} or AMEGO \cite{AMEGO}.

We perform a parametric study of the estimator from which we extract a
heuristic analytical description of the relative uncertainty of the
momentum measurement.
A good representation of these data is obtained with the following expression
\cite{Frosini:2017ftq}: 
\begin{equation}\label{eq:sigma:sur:p:parametrique}
 \gfrac{\sigma_p}{p} \approx \gfrac{1}{\sqrt{2 N}}
 \sqrt[4]{
1 + 256
 \left( \gfrac{p}{p_0} \right)^{4/3}
 \left( \gfrac{\sigma^2 X_0}{N \Delta x ~ l^2} \right)^{2/3}
} ,
\end{equation}

$\sigma$ is the single-measurement spatial RMS precision and $N$ the
number of measurements.
The high-momentum asymptote is found to have the same functional
dependence on the detector parameters as was obtained with the
back-of-the-envelope calculation of \cite{Bernard:2012uf},

The method is also of interest for the high-precision measurement of
muon momentum in large, finely-instrumented liquid argon detectors for
long-range neutrino studies such as DUNE \cite{Acciarri:2016ooe}.

It is a pleasure to acknowledge the support of the French National
Research Agency (ANR-13-BS05-0002).


\begin{thebibliography}{99}
\small
\bibitem{HARPO}
 HARPO, ``hermetic argon polarimeter'',
 {\tt 
 http://llr.in2p3.fr/$\sim$dbernard/polar/harpo-t-p.html
}
 \bibitem{Bernard:2012uf} 
 D.~Bernard,
 ``TPC in gamma-ray astronomy above pair-creation threshold,''
 Nucl.\ Instrum.\ Meth.\ A {\bf 701}, 225 (2013)
 Erratum: [Nucl.\ Instrum.\ Meth.\ A {\bf 713}, 76 (2013)].
 [arXiv:1211.1534 [astro-ph.IM]].

\bibitem{Bernard:2013jea} 
 D.~Bernard,
 ``Polarimetry of cosmic gamma-ray sources above $e^+e^-$ pair creation threshold'',
 Nucl.\ Instrum.\ Meth.\ A {\bf 729} (2013) 765.
 [arXiv:1307.3892 [astro-ph.IM]].

\bibitem{Gros:SPIE:2016}
P.~Gros {\it et al.},
``Measurement of polarisation asymmetry for gamma rays between 1.7 to 74 MeV with the HARPO TPC'',
 {\bf SPIE2016}, 9905-95,
arXiv:1606.09417 [astro-ph.IM].

\bibitem{Gros:2017wyj}
 P.~Gros {\it et al.},
 ``Performance measurement of HARPO: a Time Projection Chamber as a gamma-ray telescope and polarimeter,''
 arXiv:1706.06483 [astro-ph.IM], to appear in Astroparticle Physics.

\bibitem{E-Astrogam:2016}
 A.~De Angelis {\it et al.} [e-ASTROGAM Collaboration],
``The e-ASTROGAM mission,''
 Exper.\ Astron.\ {\bf 44} (2017) 25,
 [arXiv:1611.02232 [astro-ph.HE]].

\bibitem{AMEGO}
 J. Perkins,
 ``All-Sky Medium Energy Gamma-ray Observatory (AMEGO) - A discovery mission for the MeV gamma-ray band'',
 presented at this Symposium.

\bibitem{Takahashi:2015jza} 
 S.~Takahashi {\it et al.},
 ``GRAINE project: The first balloon-borne, emulsion gamma-ray telescope experiment,''
 PTEP {\bf 2015}, no. 4, 043H01 (2015).
 
\bibitem{Tanimori:2017ihu}
 T.~Tanimori {\it et al.},
 ``Establishment of Imaging Spectroscopy of Nuclear Gamma-Rays based on Geometrical Optics,''
 Sci.\ Rep.\ {\bf 7} (2017) 41511,
 [arXiv:1702.01483 [physics.ins-det]].

\bibitem{Moliere}
G. Molière, 
``Theorie der Streuung schneller geladener Teilchen. III. Die Vielfachstreuung von Bahnspuren unter Berüksichtigung der statistischen Kopplung'',
Zeitschrift Naturforschung A {\bf 10} (1955) 177.

\bibitem{Olive:2016xmw}
 C.~Patrignani {\it et al.} [Particle Data Group],
 ``Review of Particle Physics,''
 Chin.\ Phys.\ C {\bf 40} (2016) 100001.

\bibitem{Fruhwirth:1987fm}
 R.~Frühwirth,
 ``Application of Kalman filtering to track and vertex fitting,''
 Nucl.\ Instrum.\ Meth.\ A {\bf 262} (1987) 444.
 
\bibitem{Frosini:2017ftq}
 M.~Frosini and D.~Bernard,
 ``Charged particle tracking without magnetic field: optimal measurement of track momentum by a Bayesian analysis of the multiple measurements of deflections due to multiple scattering,''
 Nucl.\ Instrum.\ Meth.\ A 867 (2017) 182,
 arXiv:1706.05863 [physics.data-an].
 
\bibitem{Matisko:Havlena:2013}
P. Matisko and V. Havlena,
``Noise covariance estimation for Kalman filter tuning using Bayesian approach and Monte Carlo'',
Int. J. Adapt. Control Signal Process. {\bf 27} (2013) 957.

\bibitem{Acciarri:2016ooe}
 R.~Acciarri {\it et al.} [DUNE Collaboration],
``Long-Baseline Neutrino Facility (LBNF) and Deep Underground Neutrino Experiment (DUNE) : Volume 4 The DUNE Detectors at LBNF,''
 arXiv:1601.02984 [physics.ins-det].
 
\end{thebibliography}
\end{document}